\documentclass[12pt]{article}
\pdfoutput=1  
\newcommand{\TITLE}{Elastic scattering of hadrons\\without optical theorem}
\newcommand{\KEYWORDS}{elastic scattering of hadrons, optical theorem, short-ranged strong interaction,  Hilbert space, S matrix theory, unitarity, Schroedinger equation}

\usepackage{amssymb,amsmath,latexsym}
\usepackage[a4paper, twoside,
            top=2.3cm,bottom=2.3cm,
            left=2.1cm,right=2.1cm,
            bindingoffset=1.0cm
            ]{geometry}

\usepackage[sort&compress, numbers]{natbib}

\usepackage{hyperref}
\hypersetup{
    pdftex,
    bookmarks=true,         
    bookmarksnumbered=true,
    bookmarksopen=true,
    bookmarksopenlevel=0,
    bookmarksnumbered=true,
    pdftitle={\TITLE},    
    pdfauthor={J. Prochazka, V. Kundrat, M. V. Lokajicek},      
    pdfsubject={HEP-theory}, 
    pdfkeywords={\KEYWORDS}, 
}

\usepackage{cleveref}  
\newcommand{\refp}[1]{(\ref{#1})}

\crefname{equation}{Eq.}{Eqs.}
\crefname{section}{Sec.}{Secs.}
\crefname{chapter}{Chapter}{Chapters}
\crefname{table}{Table}{Tables}
\crefname{figure}{Fig.}{Figs.}
\crefname{appsec}{Appendix}{Appendices}
\crefname{appchap}{Appendix}{Appendices}

\newcommand{\ket}[1]{\left| #1 \right>} 
\newcommand{\bra}[1]{\left< #1 \right|} 
\newcommand{\braket}[2]{\left< #1 \vphantom{#2} \right|\left. #2 \vphantom{#1} \right>} 

\renewcommand\Re{\operatorname{Re}}
\renewcommand\Im{\operatorname{Im}}
\makeatletter
\def\blfootnote{\xdef\@thefnmark{}\@footnotetext}
\makeatother

\begin{document}
\begin{center}
\blfootnote{
{\hspace{-8mm}\it * Corresponding author\\}
{\it Email addresses:}  jiri.prochazka@fzu.cz (Ji\v{r}\'{\i} Proch\'{a}zka), kundrat@fzu.cz (Vojt\v{e}ch Kundr\'{a}t), lokaj@fzu.cz (Milo\v{s} V.~Lokaj\'{\i}\v{c}ek)} 
{\Large\bf \TITLE} \\[8mm]

Ji\v{r}\'{\i} Proch\'{a}zka$^{\text{*}}$, Vojt\v{e}ch Kundr\'{a}t and Milo\v{s} V.~Lokaj\'{\i}\v{c}ek
\\[4mm]
{\it   

   Institute of Physics of the AS CR, v.v.i., 18221 Prague 8, Czech Republic}\\

\end{center}
\vspace{6mm}

\noindent
{\bf Abstract}\\
All contemporary phenomenological models of elastic hadronic scattering have been based on the assumption of validity of optical theorem that was overtaken from optics. It has been stated that it may be proven in particle physics. However, it will be shown that its derivation in the framework of unitary $S$ matrix theory (which is supposed to be the most general approach in this case) has been based on several requirements that do not correspond to the actual collision characteristics of two particles. It will be shown that especially in the case of short-ranged interaction (for which the theorem is used most frequently) it cannot be applied to. The analysis of corresponding collision experiments is to be done under new basic physical assumptions. The actual progress in the description of hadronic collision processes may exist only if the distribution of different initial states will be specified on the basis of impact parameter values of colliding particles and the dependence of collision probability on this parameter will be established, without limiting corresponding conclusions by the assumption of optical theorem validity from the very beginning. \\

\noindent
{\bf keywords:} \KEYWORDS


\section{\label{sec:introduction}Introduction}
Practically all hitherto models of elastic hadronic scattering have been based on the assumption of optical theorem validity. According to this theorem the total cross section $\sigma_{\text{tot}}$ at given collision energy $\sqrt{s}$ is to be proportional to the imaginary part of elastic scattering amplitude $A_{\text{el}}$ at zero scattering angle $\theta$ (or at zero four-momentum transfer $t=-4p^2\sin^2\frac{\theta}{2}$, $p$ being three momentum of one colliding particle in center-of-mass system), i.e.,  
\begin{equation}
\sigma_{\text{tot}}(s) \propto \Im A_{\text{el}}(s,\theta=0)\,.
\label{eq:optical_theorem_prop}
\end{equation}
The complex elastic scattering amplitude $A_{\text{el}}(s,\theta)$ is obtained, e.g., with the help of Schroedinger equation (in non-relativistic case). Even the phenomenological interpretation of measured elastic differential cross section of two hadrons has been derived always on the basis of optical theorem validity; the standardly chosen parametrization of $t$-dependence (fitted from measured data) has been based in principle on the assumption of maximum frequency lying at $t=0$ (any other behavior having been practically excluded).

The given theorem used now in particle physics was overtaken from optics where it developed from the formula for refraction index (defined on the basis of wave theory of light) which contained also the influence of extinction cross section (now denoted as total cross section); see the story described by Newton \cite{Newton1976}. The formula~\refp{eq:optical_theorem_prop} has been formulated practically only on the basis of experimental refraction data without any deeper theoretical reasoning.

The first appearance of optical theorem in (quantum) theoretical description of particle scattering occurred probably in Feenberg's Ph.D. thesis and in his paper \cite{Feenberg1932} from 1932. He tried to derive the relation~\refp{eq:optical_theorem_prop} on the basis of Schroedinger equation; however, his work remained without greater attention at that time. The optical theorem was also called Bohr-Peierls-Placzek relation in the past when it was applied to nuclear reactions without any proof in \cite{Bohr1939} (in 1939) and then widely used, see again \cite{Newton1976} for historical context. 

Different attempts to prove the optical theorem theoretically in particle physics have been done mainly when the collisions of fundamental particles have started to be studied and the given theorem has been applied to also in the region of strong hadron interactions. Some of these attempts to prove it in particle physics have been interpreted as successful. However, we have demonstrated recently that fundamental discrepancy has been to exist especially if the given theorem has been applied to the elastic hadron collisions \cite{Lokajicek2009_optical_theorem,Lokajicek2014_optical_theorem}.

Some arguments used to support its validity in strong interactions have been, however, still repeated. In the following we shall attempt to provide more detailed and more systematical reasoning why this theorem can be hardly applied to in any (elastic) scattering of two hadrons. We shall start from the approach of its derivation based on unitary $S$ matrix theory which is supposed to be the most general derivation of optical theorem and which is also the most widely used framework in which (elastic) scattering of hadrons is being described (some models being applied to experimental data).

The given $S$ matrix approach has been based on the assumptions concerning the basic structure of $S$ operator acting in the Hilbert space in which the incoming and outgoing states cannot be correspondingly distinguished. We shall start, therefore, by discussing the necessary Hilbert space structure formed by Schroedinger equation solutions describing collision processes; see \cref{sec:schroedinger_hilbert}. The mentioned approaches trying to prove the validity of optical theorem in $S$ matrix theory will be then analyzed in \cref{sec:s_matrix_optical_theorem}; the main derivation approach will be reviewed and  the assumptions being in contradiction to collision characteristics for strong interaction will be then specified and commented in more details. The physical requirements concerning the Hilbert space in realistic case will be discussed in \cref{sec:phys_req_hilbert_space}.
   
The contemporary models of elastic collisions of two charged hadrons have been influenced, however, not only by the discussed optical theorem but also by not respecting the difference between scattering mechanisms of Coulomb interaction (acting at distance) and short-ranged (practically contact) strong interaction, which will be mentioned in \cref{sec:elmag_vs_strong}. More detailed physical description of the collision process may be obtained, of course, only if the statistical distribution of impact parameter values of two colliding objects corresponding to experimental conditions is taken into account and the dependence of collision characteristics on this parameter is established. The basic aspects of corresponding probabilistic model proposed recently will be briefly described in \cref{sec:impact_parameter_description}. The model shows, too, that the collision process may be interpreted on fully ontological basis.

\section{\label{sec:schroedinger_hilbert}Schroedinger equation and corresponding Hilbert space}
Time evolution of microscopic processes is being described with the help of the Schroedinger (linear differential) equation
\begin{equation}
i\hbar\frac{\partial\psi(x,\tau)}{\partial \tau} = \hat{H} \psi(x,\tau)
\label{eq:schr_equ}
\end{equation}
where Hamiltonian operator $\hat{H}$ is given by
\begin{equation}
\hat{H} = -\frac{\hbar^2}{2m}\nabla^2 + \hat{V}(x)
\end{equation}
and $\hat{V}(x)$ is corresponding potential.
Its basic solutions (represented by the product of space and time functions) may be expressed in the form
\begin{equation}
      \psi_E(x,\tau) = \lambda_E(x)e^{-\text{i}E\tau/\hbar}
\label{eq:sch_equ_partial_solution}
\end{equation}
being standardly normed to one:  $\int \text{d}x\, \psi^*_E(x,\tau)\psi_E(x,\tau)=\int \text{d}x\, \left| \lambda_E(x)\right|^2=1$ (at any time $\tau$). The function $\lambda_E(x)$ of all space coordinates ($x$) corresponds to the Hamiltonian eigenfunctions at energy $E$
\begin{equation}
 \hat{H} \lambda_E(x) = E \lambda_E(x) \label{eq:schr_eq_time_independent}.
\end{equation}
General solution $\psi(x,\tau)$ of Schroedinger equation~(\ref{eq:schr_equ}) may be then written as a superposition of solutions corresponding to individual energy values
\begin{equation}
\psi(x,\tau) =  \sum_E c_E  \psi_E(x,\tau)
\end{equation}
where $c_E$ are corresponding coefficients in a linear combination of particular solutions $\psi_E(x,\tau)$; fulfilling  $\sum_E |c_E|^2=1$.

All possible functions $\psi(x,\tau)$ of space coordinates at different $\tau$ values form a complete Hilbert space. Schroedinger defined then expected values $Q(\tau)$ of physical quantities
\begin{equation}
   Q(\tau) = \int\psi^*(x,\tau)\,\hat{Q}\,\psi(x,\tau)dx    
\label{eq:expected_value}
\end{equation}
corresponding to classical quantities. It was shown originally for inertial motion only; however, it holds practically generally. Only the set of Schroedinger solutions is smaller due to discrete quantum states in closed systems. It was shown by Ioannidou \cite{Ioannidou1982} and Hoyer \cite{Hoyer2002} that the Schroedinger equation might be derived for statistical combination of Hamilton equation solutions (or be at least equivalent to these solutions) if their whole set was limited by a rather weak condition; see also \cite{Lokajicek2012_intech,Lokajicek2007}.

A $\tau$-dependent solution $\psi_E(x,\tau)$ of Schroedinger equation (the set of vectors in the corresponding Hilbert space corresponding to different values of $\tau$) represents the evolution of motion as an open trajectory in the case of continuous energy spectrum or as a closed trajectory for discrete energy values. Each physical quantity $Q(\tau)$ is then represented by associated operator $\hat{Q}$ acting in the given Hilbert space. 

Any vector $\psi_E(x,\tau)$ represents instantaneous state belonging to two opposite momentum directions. To distinguish these two different cases the total Hilbert space (in the case of elastic collisions) must consist of two mutually orthogonal subspaces each being spanned on the basis of Hamiltonian eigenfunctions $\lambda_E(x)$ as it has been shown already many years ago by Lax and Phillips \cite{Lax1967,Lax1976} and independently derived also by Alda et al.~\cite{Alda1974} from the requirement of exponential (purely probabilistic) decay law of unstable particles. Only in such an extended Hilbert space the collision processes of two particles may be correspondingly described. 
 The transition from one subspace to another may be then given by the evolution operator 
\begin{equation}
     \hat{U}_{ev}(\tau) \; =\; e^{-\text{i}\hat{H}\tau/\hbar}; 
\label{eq:ev_op}
\end{equation}
 the opposite evolution corresponding to negative values of $\tau$. It holds then
\begin{equation}
     \psi_E(x,\tau) \;=\; \hat{U}_{ev}(\tau)\psi_E(x,0) \,.
 \end{equation}
If $\psi_E(x,0)$  represents the state corresponding to the shortest distance between two colliding particles then the states for time $\tau>0$ belong  to the subspace of outgoing particles and for time $\tau<0$ to that of incoming states.
 
The given Hilbert structure has been, however, excluded by Bohr in 1927 \cite{bohr1928} who asked for the Hilbert space of any physical system to be spanned always on one basis of Hamiltonian eigenvectors. It has caused that the earlier physical interpretation of Schroedinger equation solutions has been fundamentally deformed as any description of continuous time evolution has been practically excluded. Moreover, the given model has required the existence of immediate interaction between very distant particles, which was shown and criticized by Einstein in 1935 with the help of special coincidence Gedankenexperiment. The physical scientific community preferred, however, Bohr's approach (in the region of microscopic processes). 
    
Later both the alternatives were admitted and discussed. Bohr's alternative was, however, supported again on the basis of the fact that Bell's inequality (derived in 1964 for the coincidence experiment more specified than that of Einstein) was violated in the corresponding experiment including spin measurement and performed by Aspect et al.~in 1982 \cite{Aspect1982}. It has been shown only recently that Bell's inequality was based always on an assumption that did not hold in the given more specified experiment (but only in that proposed originally by Einstein); see, e.g., \cite{Lokajicek2012_intech}. Consequently, Einstein has been fully right in the given controversy with Bohr and the Hilbert space must always consist at least of two mutually orthogonal subspaces as explained in the preceding. All necessary details may be found in \cite{Lokajicek2012_intech,Lokajicek2012_scripta,Lokajicek2013_intech} and \cite{Lokajicek2013_bell}.

\section{\label{sec:s_matrix_optical_theorem}  S matrix theory and optical theorem}

In the  region of strong interactions the decisive study of elastic processes has concerned two-proton collisions where the experimental data especially for small scattering angles have represented the combination of Coulomb and hadronic interactions. The ratio of these two interactions has always being determined on the basis of some theoretical predictions. However, the contemporary approaches (in both the interaction kinds) have started often from some assumptions that have not corresponded to actual experimental arrangement or to differences in divers interaction kinds as it will be shown in the following. 

As to the Coulomb interaction it has been assumed that the corresponding elastic differential cross section has risen to infinity for very small scattering angles, which has followed from the fact that the zero scattering angle should be obtained at infinite distance (i.e., at infinite impact parameter). However, the measured region of scattering angles corresponds to impact parameter values of less than micrometers, which has not been respected in the usual formula that has been used for interpretation of the Coulomb part of measured data. In addition to, a part of measurable elastic collisions may be caused by multiple Coulomb scattering according to experimental conditions (target density).

Similar criticism concerns, of course, the assumed behavior of strong interactions in the same region. Here, the validity of optical theorem given by \cref{eq:optical_theorem_prop} has been assumed practically in all theoretical as well as experimental studies which strongly influenced $t$-dependence of hadronic elastic differential cross section in neighborhood of $t=0$. 

The optical theorem has been overtaken from optics without having been proved in the past. It will be shown that also all contemporary attempts to prove its validity in particle physics have been based on assumptions that are not surely valid in the case of strong interaction. As it has been already mentioned the main attempt to derive the optical theorem in particle physics was done in the framework of $S$ matrix theory in which some important assumptions concerned the structure of corresponding Hilbert space as well as of $S$ matrix itself. The collision process has been described with the help of $S$ operator acting in Hilbert space spanned on all possible states and defined in principle phenomenologically. The $S$ matrix was introduced for the first time by Wheeler \cite{Wheeler1937} in 1937 who concluded that it should be unitary. Heisenberg \cite{Heisenberg1943_one,Heisenberg1943_two} invented the $S$ matrix for description of scattering processes independently in 1943 as the communication between German and Western scientists was disrupted due to the second world war, see \cite{Newton1976} for some more historical comments. He also stated that the $S$ matrix should be unitary and, at the difference to Wheeler, also tried to derive the optical theorem in this theoretical framework.  

In the following we shall review some basic formulas which have been used in different attempts to prove optical theorem based on $S$ matrix, using the conventions from \cite{Barone2002};  for more details and comments see also, e.g., \cite{Galindo1990_vol2,Pilkuhn1967,Pilkuhn1979,Merzbacher1961,Messiah1961_vol2} or any other textbook covering $S$ matrix and optical theorem. Problematic steps will be identified and discussed in more detail in \cref{sec:OT_assumptions}.
 
\subsection{\label{sec:s_matrix_ot_derivation} Main approach of deriving optical theorem} 

The $S$ operator has been proposed  to transform an initial two-particle state $\ket{i}$ characterized by momentum (energy) of the colliding particles directly to a possible final state $\ket{f}$:   
\begin{equation}
    \ket{f} = S \ket{i} \,.
\label{eq:S_operator}
\end{equation}
The matrix elements of this operator have been assumed to determine the probability of corresponding transitions (scattering)  
\begin{equation}
   P_{i\rightarrow f} = \left|\bra{f}S\ket{i}\right|^2.    
\label{eq:p_if}
\end{equation}
The conservation of probabilities in the given approach has been written in the form
\begin{equation}
   \sum_f P_{i\rightarrow f} = 1    
\label{eq:sum_p_if}
\end{equation}
where it has been summed over all possible final states. The given $S$ operator has been required to fulfill the condition of unitarity 
\begin{equation}
     S^+S = S\,S^+=\,I \;.
\label{eq:s_unitary}
\end{equation}
Practically in all approaches attempting to derive optical theorem the transition operator $T$ has been defined in the form
\begin{equation}     
     S = I + \text{i}T   
\label{eq:s_matrix}
\end{equation}
where the introduction of identity operator $I$ has been based on the assumption that one can  ''separate'' non-interacting part ($S=I$) from the case of an interaction. The matrix elements of $S$ operator  have been then given as
\begin{equation}
S_{if} \equiv \bra{f} S \ket{i}  = \braket{f}{i} + \text{i} \bra{f} T \ket{i} \, .
\end{equation}
Scattering amplitude $A(i \rightarrow f)$ has been then defined (using matrix elements of $T$) as 
\begin{equation}
T_{if}   \equiv  \bra{f} T \ket{i} = (2 \pi)^4 \delta^4(p_f - p_i) A(i \rightarrow f)
\label{eq:def_ampl}
\end{equation}
where $\delta^4(p_f - p_i)$ has ensured four-momentum conservation. It has also meant that
\begin{equation}
T^*_{fi} \equiv \bra{f} T^+ \ket{i} = (2 \pi)^4 \delta^4(p_i - p_f) A^*(f \rightarrow i).
\label{eq:def_ampl_conjugate}
\end{equation}

Partial cross sections (corresponding to different collision channels) in this $S$ matrix approach have been defined using the scattering amplitude ($T$ matrix elements) as
\begin{equation}
\text{d}\sigma = \frac{1}{\phi} \left|A(i \rightarrow f_n)\right|^2 \text{d}\Pi_n.
\label{eq:ampl_dcs}
\end{equation}
Incident flux $\phi$ in \cref{eq:ampl_dcs} has been defined as
\begin{equation}
\phi = 2 E_1 \, 2E_2 |\boldsymbol{v}_1 - \boldsymbol{v}_2 |
\end{equation}
where $\boldsymbol{v}_1$ and $\boldsymbol{v}_2$ has been the three-dimensional velocities of the colliding particles 1 and 2, respectively, i.e., $\boldsymbol{v}_1 = \frac{\boldsymbol{p}_1}{E_1}$ and $\boldsymbol{v}_2 = \frac{\boldsymbol{p}_2}{E_2}$; $E_1$ and $E_2$ being energies of the incoming particles with masses $m_1$ and $m_2$. The incident flux has been then simplified to the form
\begin{equation}
\phi = 2 \sqrt{ \lambda(s,m_1^2, m_2^2)} 
\end{equation}
where the $\lambda$ function has been defined as
\begin{equation}
\lambda(x,y,z) = x^2 + y^2 + z^2  - 2 xy - 2 yz - 2 xz, \;\;\; \Rightarrow \lambda(x,y,y) = x^2 - 4 xy.
\end{equation}
The factor $\text{d}\Pi_n$ stands for the Lorentz-invariant phase space 
\begin{equation}
\text{d}\Pi_n = \prod_{j=1}^n \frac{\text{d}^3 \boldsymbol{p}^\prime_j}{(2\pi)^3 2E^\prime_j} (2\pi)^4 \delta^4(p_1+p_2 - \sum_{j=1}^n p^\prime_j)
\label{eq:lips}
\end{equation}
for $n$ particles in the final state (corresponding values denoted by prime).

In the case of elastic differential cross section as a function of energy $\sqrt{s}$ and four-momentum transfer $t$ \cref{eq:ampl_dcs} has been simplified to
\begin{equation}
\frac{\text{d}\sigma_{\text{el}}}{\text{d}t}(s,t) = \frac{1}{16 \pi \lambda(s,m_1^2,m_2^2)} \left|A_{\text{el}}(s,t)\right|^2 \, .
\label{eq:ampl_dcs_el}
\end{equation}
Integrating \cref{eq:ampl_dcs} over all ($n$-particle) final states and summing over $n$ (i.e., integrating over all possible final states) it has been obtained for total cross section 
\begin{equation}
\sigma_{\text{tot}} = \frac{1}{\phi} \sum_n \int \text{d}\Pi_n \left|A(i \rightarrow f_n)\right|^2 \, .
\label{eq:sigma_tot_T_matrix}
\end{equation}

It has then followed from the unitarity of $S$ operator given by \cref{eq:s_unitary} for the $T$ operator defined by \cref{eq:s_matrix} 
\begin{align}
(I-\text{i}T^+)(I+\text{i}T) &= I\;, \\
\text{i}(T^+-T) &= T^+ T \,.
\label{eq:T_manipulations_0}
\end{align}
\cref{eq:T_manipulations_0} has been then rewritten in matrix form as
\begin{equation}
\text{i}\bra{f} T^+ - T \ket{i} = \sum_{\{n\}} \bra{f} T^+ \ket{n} \bra{n} T \ket{i}  \, .
\label{eq:T_manipulations_1}
\end{equation}
where ''intermediate'' states $\ket{n}$ have represented the complete set of final states and the l.h.s.~has been simplified to
\begin{equation}
\text{i}(T^*_{fi} - T_{if}) =  \text{i}((\Re T_{fi} - \text{i} \Im T_{fi}) - (\Re T_{if} + \text{i} \Im T_{if})) = 2 \Im T_{if}.
\label{eq:T_manipulations_2}
\end{equation}
Last equality in \cref{eq:T_manipulations_2} holds if and only if $T$ is symmetric, i.e., if $T_{if} = T_{fi}$. \cref{eq:T_manipulations_1,eq:T_manipulations_2} have implied then
\begin{equation}
2 \Im T_{if} = \sum_{\{n\}} T^*_{fn} T_{in} \, .
\label{eq:T_linear_quadratic_condition}
\end{equation}

It has been then deduced on the basis of \cref{eq:def_ampl,eq:T_linear_quadratic_condition} that in the case of initial and final states being identical (i.e., if $f = i$ which in $2 \rightarrow 2$ particle collision has been identified with the case of elastic collisions corresponding to zero scattering angle, $t=0$) one may obtain
\begin{equation}
2 \Im A_{\text{el}}(s,t=0) = \sum_n \int \text{d}\Pi_n  \left|A(i \rightarrow n)\right|^2 \,.
\label{eq:ampl_unitarity}
\end{equation}
Finally, it has been concluded (comparing \cref{eq:sigma_tot_T_matrix,eq:ampl_unitarity}) that it should hold
\begin{equation}
\sigma_{\text{tot}} = \frac{2}{\phi} \Im A_{\text{el}}(s,t\!=\!0) \, ,
\label{eq:OT}
\end{equation}
i.e., imaginary part of elastic scattering amplitude at zero scattering angle ($t=0$) should be proportional to total cross section - the relation commonly known as optical theorem, see also \cref{eq:optical_theorem_prop}.

\subsection{\label{sec:OT_assumptions} Basic assumptions and physical reality}
In the preceding subsection we have reviewed the main approach of deriving optical theorem in particle physics with the help of (relativistic) $S$ matrix theory. This subsection is devoted to discussion of various assumptions which were used in the corresponding approach.

\begin{enumerate}
\item{The initial two-particle state $\ket{i}$ has been characterized by the momenta and energy of colliding particles but not by their impact parameter. Only one ''average'' two-particle initial state has been taken into account in the given $S$ matrix approach (some related discussion may be found in Chap.~8 of \cite{Galindo1990_vol2}; in other textbooks this aspect has not been usually  mentioned at all). The orientation of angular momentum vectors has not been distinguished in the given Hilbert space, either. The probability~\refp{eq:p_if} and cross section~\refp{eq:ampl_dcs} of a given transition $i \rightarrow f$ from one initial to a final state correspond, therefore, only to some average values.}

\item{In all attempts the approach to derive the optical theorem in $S$ matrix theory has been based on \cref{eq:T_linear_quadratic_condition} for special case $\ket{f}=\ket{i}$ denoted as ''non-interacting'' case. It has been admitted that the average initial state $\ket{i}$ may represent a final state, too. The non-interacting case $\ket{f}=\ket{i}$ has been then artificially identified with the (unmeasurable) limit value of elastic scattering at $t=0$ in \cref{eq:ampl_unitarity}.

However, in the case of short-ranged strong interaction very many events at higher impact parameter values may exist when the incoming particles do not interact at all. These non-interacting events correspond to zero ''scattering'' angle ($t=0$) but their frequency may be very much higher than in the case corresponding to elastic scattering in unmeasurable limit point $t\rightarrow0$. One should expect completely different transition probabilities for these two cases. Moreover, the measured frequency of non-interacting events may depend on experiment arrangement (target density). This non-interacting transition in the case of very short-ranged hadronic interaction may have probability close to $\;1\;$ and the elastic scattering in limit point $t\rightarrow0$ (and in all interacting events) should have very small probability (close to zero) as the majority of colliding particles (at higher impact parameters) may simply pass without any interaction. The optical theorem has been obtained by identifying (interchanging) quite arbitrarily these two cases (limit value of elastic scattering and all non-interacting events).

Two different cases should be carefully distinguished: $\ket{f}=\ket{i}$ and $\ket{f}\neq\ket{i}$, i.e., non-interacting and interacting.
In the interacting case $\ket{f}\neq\ket{i}$ the corresponding transition (scattering) probability given by \cref{eq:p_if} is
\begin{equation}
P_{i \rightarrow f} \equiv \left|\bra{f} S \ket{i}\right|^2 = \left|\bra{f} S - I \ket{i}\right|^2 = \left|\bra{f} T \ket{i}\right|^2.
\end{equation}
The probability (cross sections) in this case may be calculated using either $S$ or $T$ matrix elements.
However, in the non-interacting case $\ket{f}=\ket{i}$ the scattering operator has been taken as identity operator (compare with \cref{eq:S_operator}) which implies
\begin{equation}
P_{i \rightarrow i} \equiv \left|\bra{i} S \ket{i}\right|^2 = \left|\bra{i} I \ket{i}\right|^2 \neq \left|\bra{i} S - I \ket{i}\right|^2 = \left|\bra{i} T \ket{i}\right|^2 \, .
\end{equation}
The diagonal matrix element $T_{\,ii}\,$ does not, therefore, represent the corresponding probability; $S_{ii}$ has to be used instead of $T_{\,ii}$, which has not been taken into account in the attempts to derive optical theorem. The differential cross section has always been calculated on the basis of the $T$ matrix ($S-I$ matrix), see \cref{eq:ampl_dcs}.}


\item{\cref{eq:ampl_dcs} contains also the factor $\text{d}\Pi_n$ given by \cref{eq:lips} which has been introduced (see, e.g., \cite{Pilkuhn1967}) without sufficient reasoning. The factor does not distinguish between the interacting and the non-interacting case in the processes $2 \rightarrow 2$ ($n=2$), even though these two cases are completely different as it has been discussed. The introduction of this additional factor for each $n$-particle final state $\ket{f_n}$ has changed the original probabilistic interpretation of $T$ matrix elements, too.}

\item{The optical theorem given by \cref{eq:OT} has been obtained comparing \cref{eq:ampl_unitarity,eq:sigma_tot_T_matrix}. The integration in \cref{eq:ampl_unitarity} has been done over the complete set of all considered states (all considered transitions); therefore no state has been excluded. The integration in \cref{eq:sigma_tot_T_matrix} has been taken also over all final states (again all transitions); however, the non-interacting events $i \rightarrow i$ mentioned above should be excluded here if the l.h.s.~is to correspond to total cross section of two-particle collisions. 

If one excludes the non-interacting case in the integration in \cref{eq:sigma_tot_T_matrix} then it may be possible to obtain the following relation

\begin{equation}
\sigma_{\text{tot}} = \frac{2}{\phi} \Im A_{\text{el}}(s,t\!=\!0) - \frac{1}{\phi} \int \text{d}\Pi_2 \left|A(i \rightarrow i)\right|^2 
\label{eq:OT_improved}
\end{equation}
instead of the optical theorem given by \cref{eq:OT}. It means that the second term on r.h.s.~of \cref{eq:OT_improved} has been wrongly put to zero (neglected) to obtain the optical theorem. The absolute value of the second term might be much bigger than the first one as in the case of short-ranged hadronic interaction high fraction of colliding particles cannot interact at higher impact parameters at all. The total cross section given by \cref{eq:OT_improved} might be even negative which indicates the fact that the whole derivation of optical theorem has been based on some unphysical assumptions. 
}
\item{The $S$ matrix has been assumed to be unitary (also inverse matrix exists) or even symmetric due to \cref{eq:T_linear_quadratic_condition} \footnote{The requirement of $S$ being symmetric implies $P_{i \rightarrow f} = P_{f \rightarrow i}$ and is related to often assumed ''time-reversal invariance'' of corresponding interactions (transitions $i \leftrightarrow f$).}, which means in both the cases that each initial state should represent also a final state and vice versa; being the consequence of the Hilbert space formed only by single Hilbert subspace. Galindo and Pascual \cite{Galindo1990_vol2} (page 17) have pointed out that ''$S$ is unitary if and only if $H_{\text{in}} = H_{\text{out}}$'', i.e., if and only if the Hilbert space formed by initial states is the same as that of final (outgoing) ones. They have also pointed out (in agreement with \cite{Hunziker1968}) that ''the unitarity of $S$ is not simply a consequence of probability conservation'' (contrary to the widely used statements). Already the assumption of $S$ matrix being unitary has more consequences than it has been generally assumed. E.g., the optical theorem has excluded the possibility of hadronic elastic differential cross section being equal to zero at $t=0$, even if such possibility may be physical and should be admitted in further analysis of elastic processes. It has not been explained in derivation of optical theorem, either, why on one hand all ($n$-particle) final states may represent also initial ones (requirement of one Hilbert subspace) but on the other hand all cross sections (transitions) are calculated only from one average two-particle initial state. The initial states in collision experiments (with colliding beams or fixed target) correspond always to two-particle states typically at one quite well defined value of collision energy. Therefore, admitting $n$-particle initial states ($n>2$) in description of experimental data to derive some properties of colliding particles is unphysical. This represents another inconsistency in the usual derivation of optical theorem in particle physics.}
\end{enumerate}

The optical theorem has been introduced on the basis of the $S$ matrix theory in a quite abstract manner without explicit form of the $S$ operator. The $S$ matrix elements, or rather the scattering amplitudes $A(i \rightarrow f)$, have been determined either from some theoretical calculations (e.g., introducing the Hamiltonian of a given system and using Schroedinger equation, or similarly in relativity) or experimentally (phenomenologically). In the former case no reliable theory of elastic hadron collisions is available (some perturbation approaches, used more frequently for inelastic channels, are inapplicable in this case). The elastic scattering amplitude determined from experimental data (also needed for optical theorem derivation) has been quite arbitrarily parameterized on the basis of optical theorem validity and then fitted to measured elastic differential cross section of two hadrons. Contemporary models of elastic hadronic scattering may be, however, hardly very helpful (see, e.g., \cite{Kaspar2011}).

The assumptions under which the optical theorem has been derived have not been practically tested experimentally. All experimental ''tests'' of optical theorem have corresponded basically only to comparison of values of total hadronic cross sections determined with and without using optical theorem, see, e.g., \cite{Eberhard1972,Eberhard1974} ($\pi$p scattering) or \cite{Amaldi1976,totem6} (pp scattering). However, even if both the values may be found to be similar one cannot conclude that the optical theorem is valid (contrary to statements in \cite{Eberhard1972,Eberhard1974}). The experimental testing is more delicate as the optical theorem is always accompanied also by some other strong assumptions. 

The value of total hadronic cross section determined on the basis of optical theorem has been, e.g., strongly influenced by the choice of elastic hadronic scattering amplitude in the close neighborhood of $t=0$, i.e., in the region which has not been experimentally reachable (especially, in the  presence of Coulomb interaction that is much stronger and dominant here). It has always been assumed from the very beginning that the modulus of the elastic hadronic scattering amplitude has been (quasi-)exponential in this region, decreasing with rising scattering angle. The corresponding parametrization has been influenced decisively by assumed validity of optical theorem.

As it has been mentioned in the introduction to \cref{sec:s_matrix_optical_theorem} it has been also the singularity of standardly used Coulomb amplitude at zero scattering angle corresponding to infinite impact parameter which has not corresponded to the experimental conditions since the values of impact parameter (in any elastic experiment) are always much lower. It is rather the effect of a greater number of Coulomb interactions (with different more distant targets) in individual events, which should correspond to measured scattering angles around zero. The Coulomb interaction contributes consequently much more to greater scattering angles than it has been assumed usually in hadronic collisions. The contemporary analysis of collision processes is also probably strongly influenced by not distinguishing the effect of long-ranged Coulomb and of short-ranged strong interactions even though they are very different as it will be discussed in \cref{sec:elmag_vs_strong}. 


The attempt to derive the optical theorem from non-relativistic Schroedinger equation may be found, e.g., in \cite{LandauLifshitz_QM_1965} (1965), where also explicit $S$ operator has been shown. Therefore, it might be possible to identify all the included problems mentioned above also in this case. Some other approaches trying to prove optical theorem have been summarized in \cite{Barone2002}. All contemporary relativistic as well as non-relativistic attempts to prove optical theorem in particle physics are, however, quite similar and contain all main problems discussed above, even though they are not (explicitly) formulated on the basis of the unitary $S$ matrix.  

In any case one can conclude from the preceding that the derivation of optical theorem is unacceptable especially for short-ranged strong interactions for the reasons introduced above. The whole description of the physical process may be strongly deformed if the optical theorem validity has been applied to as a basic assumption. Consequently, some new descriptions of collision processes without optical theorem constraint be looked for. 
\\

\section{\label{sec:phys_req_hilbert_space} Hilbert space structure and S operator corresponding to physical reality}

The basic problem in derivation of the optical theorem consists in phenomenological definition of $S$ matrix and corresponding Hilbert space where incoming and outgoing (initial and final) collision states belong to one Hilbert subspace (are spanned on one common set of basic orthogonal vectors) and may be represented as mutual superpositions. It may be seen from \cref{sec:schroedinger_hilbert} that the given states represented by solutions of Schroedinger equation may be distinguished only if the $S$ operator is defined as acting in Hilbert space consisting of two mutually orthogonal Hilbert subspaces, one containing different initial states and the other containing corresponding final ones
\begin{equation}
 H\,=\,H_i\oplus H_f  
\end{equation}
where the incoming or outgoing states are represented always by different vectors;  at least the direction of mutual evolution being always distinguished. The incoming two-particle states $\ket{i}$ are to be distinguished then not only by collision energy (momenta of colliding particles) but also at least by impact parameter values usually distributed statistically. The introduction of the impact parameter is also necessary for definition of angular momentum, the quantity being conserved (in addition to energy) during evolution of any physical system.

Different kinds of outgoing (final) states need to be carefully distinguished, too. The subspace $H_f$ is to be divided at least into three orthogonal subspaces
\begin{equation}
 H_f\,=\,H_{f_\text{el}}\oplus H_{f_\text{unsc}} \oplus H_{f_\text{inel}} \, .
\end{equation}
The subspace $H_{f_\text{el}}$ represents subspace of elastic final states distinguished by scattering angle or equivalently. Hadronic collisions at a given impact parameter value may lead to an interval of scattering angles (according to instantaneous space orientations of colliding protons that should be regarded as non-spherical). It is natural to assume in both the Coulomb and hadronic cases that the average scattering angle increases with decreasing impact parameter value. However, in short-ranged strong interaction only a very small part of events (corresponding to impact parameters less than several femtometers) will contribute to total hadronic cross section while a much greater part of events (corresponding to impact parameters of greater values) will pass without any strong interaction. 

$H_{f_\text{unsc}}$ corresponds then to the subspace of non-interacting events differing decisively from final elastic states (including the state corresponding to zero limit of scattering angle); they may be distinguished similarly as initial states (e.g., by corresponding impact parameter). If we start from the ontological interpretation of collision processes (see the end of \cref{sec:schroedinger_hilbert}) it is necessary to expect that the Coulomb and strong interaction will behave very differently. The subspace $H_{f_\text{unsc}}$ is to be empty in the case of long-ranged Coulomb interaction and non-empty in the hadronic one.

The inelastic processes are then represented by transitions to $H_{f_{\text{inel}}}$. The final states may be represented by corresponding subspace where different numbers of particles may be produced by the decay of excited unstable objects having been created in the collision of incoming particles. The subspace $H_{f_\text{inel}}$ may be regarded as the sum of other orthogonal subspaces representing different inelastic processes, even if the initial (incoming) states corresponding to a greater number of final particles cannot be realized and do not exist. The actual structure of subspaces $H_{f_\text{inel}}$ is, therefore, open for future theoretical analysis. 

The $S$ operator should then define the transition probabilities of an initial state to some state belonging to one of three divers final subspaces; no opposite transitions existing. It means that $S$ operator cannot be unitary, see also \cref{sec:OT_assumptions}. If more different initial states than one (average) 
two-particle state are taken into account then the conservation of probability should be written as (generalization of \cref{eq:sum_p_if})
\begin {equation}
            \sum_i \sum_f   w_i P_{i\rightarrow f} = \sum_i \sum_f w_i |\bra{f}S\ket{i}|^2  =1     
\label{eq:prob_conservation_new}
\end{equation} 
where $w_i$ represents the weight of a given initial state (corresponding to, e.g., distribution of impact parameters - experimental conditions); holding $\sum_i w_i = 1$. It should also hold
\begin {equation}
      |\bra{i}S\ket{f}|^2=0 \;\; \forall i, \forall f 
\end{equation} 
i.e., time irreversibility of the scattering process (in the framework of corresponding Hilbert space).

As to the optical theorem applied to particle physics some problems have been pointed out  already by Kupczynski at least since 1973; see, e.g., \cite{Kupczynski1973,Kupczynski1974,Kupczynski1977,Kupczynski1987} and also recent paper \cite{Kupczynski2013}. In the last paper he has also considered existence of some mutually orthogonal Hilbert subspaces (see also \cite{Kupczynski1977}). However, there is a difference between both the approaches. In our approach incoming and outgoing states belong to mutually orthogonal subspaces as it is required if incoming and outgoing particles are to be represented always by different vectors in corresponding Hilbert space (see \cref{sec:schroedinger_hilbert}) while in the system assumed in \cite{Kupczynski2013} each subspace should be divided into two orthogonal subspaces, yet. According to Kupczynski one may have unitary $S$ matrix without optical theorem while we have provided several arguments also against unitarity of the $S$ matrix, see \cref{sec:OT_assumptions}.  In both the approaches (of Kupczynski and of ours) it has been, however, demonstrated that the optical theorem is to lead to great deviations from physical reality especially in the case of strong interaction. Further progress in hadron physics may be hardly reached with making use of optical theorem and ignoring its assumptions and far-reaching consequences.

\section{\label{sec:elmag_vs_strong}Difference of electromagnetic and strong interaction mechanisms}
The determination of elastic differential cross section of two charged hadrons from the measured data may be significantly influenced by one factor more. There is important difference in the mechanism of the Coulomb and strong interactions. While the Coulomb interaction acts at greater distances between colliding hadrons and may be described with the help of corresponding potential the strong interaction should be interpreted rather as contact one and the corresponding description of its effect should be looked for. 
 
Consequently, the probability of strong collision at actual (initial) impact parameter value may be influenced significantly by the continuous effect of the Coulomb interaction at greater distances before a proper (contact) collision may happen. The actual minimum mutual distance (in proper collision moment) of strongly colliding particles may be significantly influenced by distant Coulomb interaction especially at lower collision energy values. It means that, e.g., for pp collisions the established value of total hadronic elastic cross section may be lower than it corresponds to reality. The given influence may be rather different at divers collision energy values, the frequency of strong interaction events corresponding to initial impact parameter being decreased or increased according to charges of colliding particles (differently in dependence on collision energy).
  
However, until now the difference between the long-ranged Coulomb and short-ranged (contact) hadronic interaction (if they act simultaneously) has not been taken into account. The analysis of (experimental) results published earlier should be, therefore, examined also under these new conditions; see the next section.

\section{\label{sec:impact_parameter_description} New probabilistic model of collision processes} 

The hitherto models of elastic nucleon collisions have been in principle phenomenological, looking mainly for a simple description of scattering characteristics. However, when one is to understand better corresponding physical mechanism the distribution of at least some other initial state characteristics must be taken into account (in addition to particle momenta). It is mainly the distribution of initial states distinguished also by different impact parameter values corresponding to experimental conditions which should be taken into account in any description of (elastic) collision processes (see \cref{eq:prob_conservation_new}).

Such a description trying to take more realistic behavior of particle collisions in the impact parameter space into account has been proposed by us in 1994 \cite{Kundrat1994_unpolarized}, see also \cite{Kundrat2002,Kaspar2011}. However, even if it has been possible to study some new characteristics of elastic collisions (e.g., mean values of impact parameter corresponding to total, elastic and inelastic collisions) the deformation caused by assuming the validity of commonly accepted optical theorem has remained until now. Systematic arguments against the validity of optical theorem in particle physics has been discovered fully only recently.

If the limitation given by optical theorem is not applied to then a quite new approach may be used for description of elastic collision processes. The corresponding collision model of charged hadrons has been recently proposed by us in \cite{Lokajicek2009_model_prob,Lokajicek2013_intech}. Starting from the ontological interpretation of colliding objects and assuming that these objects are not fully spherical (differently oriented in space) one should expect that the probability of collision processes will depend mainly on the values of mutual impact parameter $b$ (distributed according to experimental arrangement). It may be then written for the probability of elastic hadronic collisions at given impact parameter $b$
\begin{equation}
   P^{\text{el}}(b) \;=\; P^{\text{tot}}(b)\,P^{\text{rat}}(b) 
\end{equation}
where $P^{\text{tot}}(b)$ is the probability of all possible hadronic (elastic or inelastic) collision processes and $P^{\text{rat}}(b)$ is the mutual ratio of elastic probability to total one at corresponding value of impact parameter $b$. 

In the case of short-ranged (contact) strong interactions one may expect further on the basis of ontological realistic approach that elastic collisions will be mainly peripheral. The functions $P^{\text{tot}}(b)$ and $P^{\text{rat}}(b)$ may be then assumed to be monotonous functions of $b$: the first one diminishing with rising $b$ and the other increasing in the same interval of $b$. If one admits that a proton may exist in some internal states differing at least very slightly in their dimensions then both the monotonous functions (for individual proton collision channels) may be determined from corresponding measured elastic differential cross section. 

These probabilities as the functions of impact parameter provide much more information than probabilities $P_{i\rightarrow f}$ corresponding to just one average initial state which were discussed in \cref{sec:s_matrix_optical_theorem}. This also means that the new approach may provide deeper understanding concerning the structure and interactions of colliding particles.

The new collision model has been applied (in its preliminary form) to experimental data represented by measured elastic proton-proton differential cross section at energy of 52.8~GeV. It has been possible to demonstrate explicitly that new possibilities of fundamental particle research have been opened on its basis; see \cite{Lokajicek2013_intech} for more details.

The influence of distant Coulomb interaction has not yet been taken into account in this preliminary analysis.  The more detailed analysis of proton-proton collisions under all new conditions is being prepared. 

\section{\label{sec:conclusion}Conclusion}

All attempts of proving optical theorem in particle physics has started practically from $S$ matrix approach based on the assumption that the corresponding Hilbert space ($S$ operator acting in) has consisted of the superpositions of incoming and outgoing states as it was required by Bohr. To respect the realistic (ontological) characteristics of (elastic) collision processes the initial and final states are to be represented in two mutually orthogonal subspaces of the Hilbert space formed by the solutions of corresponding Schroedinger equation (in non-relativistic case) as it was discussed in \cref{sec:schroedinger_hilbert,sec:phys_req_hilbert_space}. Only then the difference between initial and final states may be fully respected; the statistical distribution of impact parameter values in initial states and scattering angles of outgoing particles in final states being distinguished in corresponding different subspaces.

All approaches trying to prove optical theorem in particle physics have corresponded, however, to the limited Bohr's structure of Hilbert space. It has been shown in preceding that the derivation of optical theorem in the framework of unitary $S$ matrix theory (which has been supposed to be the most general approach) has been based on several assumptions that do not correspond to collisions of two particles, see \cref{sec:s_matrix_optical_theorem}. Enormous mistakes may exist in the description of collision processes when the optical theorem has been applied to in the case of (contact) strong interaction. 

Consequently, some new description of elastic collision processes of two hadrons without optical theorem constraint should be looked for. Such description should take into account the difference of electromagnetic and strong interaction mechanisms as it was discussed in \cref{sec:elmag_vs_strong}. The probabilistic character of collisions in dependence on impact parameter and the distribution of initial states corresponding to experimental conditions should be respected, too. New elastic collision model \cite{Lokajicek2013_intech} based on these requirements has been shortly characterized in \cref{sec:impact_parameter_description}. It might open a deeper insight concerning the characteristics of hadronic collision processes and proper hadronic structure.

%

\end{document}